\documentclass[preprint,aps,floatfix,superscriptaddress]{revtex4}
\usepackage{graphicx,epsfig,amssymb,amsmath,array}
\usepackage{relsize,placeins,float}
\usepackage[dvipsnames]{xcolor}
\usepackage{blindtext}
\usepackage{subcaption}
\usepackage[english]{babel}
\usepackage[percent]{overpic}
\usepackage{ragged2e}

\begin{document}
\title{Transmitted and Storage-Dominated Resonance in Fractionally Damped Unidirectionally Coupled Duffing Oscillators}

\author{Messali Rouaida}
\email{rouaida.messali@univ-setif.dz}
\affiliation{
Intelligent Systems Laboratory, Electronics Department\\
Faculty of Technology, Ferhat Abbas Setif 1 University\\
Setif 19000, Algeria
}
\author{Mattia Coccolo}
\email{mattiatommaso.coccolo@urjc.es}
\affiliation{Nonlinear Dynamics, Chaos and Complex Systems Group, Departamento de F\'{i}sica,
Universidad Rey Juan Carlos, Tulip\'{a}n s/n, 28933 M\'{o}stoles, Madrid, Spain}

\author{Miguel A.F. Sanju\'{a}n}
\email{miguel.sanjuan@urjc.es}
\affiliation{Nonlinear Dynamics, Chaos and Complex Systems Group, Departamento de F\'{i}sica,
Universidad Rey Juan Carlos, Tulip\'{a}n s/n, 28933 M\'{o}stoles, Madrid, Spain}
\affiliation{Royal Academy of Sciences of Spain, Valverde 22, 28004 Madrid, Spain}

\date{\today}

\begin{abstract}
This paper investigates resonance transmission in two unidirectionally coupled Duffing oscillators with fractional damping, where the driver is harmonically forced and the receiver is connected through a linear coupling spring. Particular attention is paid to how fractional damping in the receiver modifies amplitude amplification, energy redistribution, and the structure of the coupled response. The numerical results reveal a clear distinction between transmitted resonance, associated with a coupling-power balance consistent with direct energy transfer through the coupling spring, and storage-dominated resonance, in which the receiver still exhibits a pronounced oscillatory response while the time-averaged coupling power becomes negative under the adopted convention. In this latter regime, fractional memory promotes temporary energy accumulation within the receiver--coupling subsystem, followed by partial release through the coupling spring without any feedback on the driver dynamics. We further show that detuning the receiver natural frequency enhances the interaction between the lower-frequency transmitted response and the higher-frequency coupled response, leading to a superposed resonance regime with increased receiver amplitude, stronger localization, and sharper response. The roles of the fractional order, coupling strength, and receiver natural frequency are systematically analyzed through frequency-response curves and parametric maps. Overall, the results show how fractional memory can be used to tune resonance transmission, energy localization, and amplified response in coupled nonlinear oscillators.
\end{abstract}

\maketitle

\section{Introduction}

Coupled nonlinear oscillators arise in a broad range of physical, biological, and engineering contexts, including mechanical vibration absorbers, electronic circuits, energy harvesting devices, neural systems, and oscillator networks \cite{Geng2024,Todri2024,Wei2017,Heltberg2021,Bullo2014}. In many such systems, one subsystem acts as a driver, while another plays the role of a receiver, so that the collective dynamics cannot be inferred directly from the isolated behavior of each component \cite{Pikovsky2001,Boccaletti2002,NayfehMook1979}. When the driver is externally forced, the coupling between the subsystems may transmit oscillatory energy and generate nontrivial resonance patterns in the receiver \cite{CoccoloSanjuanDelay,CoccoloSanjuanTR}. Understanding how this transfer takes place, and under which conditions it is enhanced or suppressed, is therefore of both theoretical and practical interest.

Resonance transmission in coupled oscillators is closely related to modal interaction, phase synchronization, and energy exchange. In the simplest picture, the receiver exhibits an amplified oscillation when the forcing acting on the driver is efficiently conveyed through the coupling spring. In the present work, the term “transmitted resonance” refers to a receiver peak associated with a coupling-power balance consistent with direct energy transfer through the coupling spring. This perspective is especially useful when the analysis is not restricted to amplitudes alone, but also includes phase relations and coupling-power diagnostics.

Memory effects can substantially reshape resonance phenomena. One well-known mechanism is time-delayed feedback, which introduces a discrete form of memory and can generate resonance enhancement, multistability, and nontrivial spectral interactions in nonlinear oscillators \cite{Stepan1989,Insperger2011}. However, many realistic media, especially viscoelastic and complex mechanical systems, are more naturally described in terms of continuous memory rather than discrete delay. Fractional calculus has become a standard framework for this purpose because fractional derivatives encode hereditary effects, long-range temporal correlations, and nonlocal dissipation in a natural way \cite{Podlubny1999,OldhamSpanier1974,Kilbas2006,Mainardi2010}. In fractional oscillators, damping does not merely reduce amplitudes; it may also shift resonance conditions, modify bandwidth, and alter how energy is redistributed over multiple oscillation cycles.

Duffing-type oscillators provide a particularly relevant setting in which to investigate such effects, since they combine nonlinearity, resonance bending, and strong sensitivity to damping and forcing parameters \cite{NayfehMook1979}. Their fractional-order counterparts have already been studied in connection with vibration control, nonlinear response modification, and memory-induced changes in resonance \cite{Coccolo2023,Coccolo2024}. Even so, the specific question of how fractional damping affects resonance transmission in coupled nonlinear oscillators remains much less explored. This gap becomes especially relevant when the receiver develops a strong oscillatory response even though the average power transfer through the coupling spring is no longer positive over the observation window.

The present paper focuses on precisely this situation. We consider two unidirectionally coupled Duffing oscillators with fractional damping, where the driver is harmonically forced and the receiver is connected to it through a linear coupling spring. Our interest is not limited to the conventional transmitted-resonance picture, but extends to a second type of response in which the receiver still displays a pronounced oscillation while the time-averaged coupling power becomes negative under the adopted convention. Within the unidirectional model studied here, this does not imply a dynamical feedback from the receiver to the driver; rather, it is consistent with the receiver--coupling subsystem releasing, on average, part of the energy previously accumulated within that part of the system through the coupling spring. This distinction motivates a classification based jointly on amplitude, phase coherence, and average coupling power.

We also examine how detuning the receiver natural frequency modifies the interaction between the lower-frequency transmitted response and the higher-frequency coupled response. In this regime, the receiver may develop a markedly enhanced oscillation that can be interpreted in connection with the resonance-transmission and resonance-superposition effects previously reported in delayed Duffing-type settings \cite{CoccoloSanjuanDelay,CoccoloSanjuanTR}. The analogy is conceptually useful, although the physical origin is different: in those earlier works, the relevant memory mechanism was discrete and delay-induced, whereas in the present study, it arises from continuous memory associated with fractional damping.

The main questions addressed in this work are therefore the following: how the receiver fractional order and the coupling strength affect resonance amplitude, sharpness, and energy localization; under which conditions the receiver develops a strong response that cannot be interpreted solely in terms of direct transmitted resonance; and how frequency detuning promotes the interaction between response branches and leads to an amplified regime. To answer these questions, we combine frequency-response analysis with phase, energy, transmissibility, and coupling-power diagnostics, and complement them with parametric maps in the relevant control planes. The results show that fractional memory can substantially reshape the pathways of energy redistribution in coupled Duffing oscillators by modifying not only the amplitude of the receiver response, but also the balance between direct transmission, temporary energy storage, and detuning-enhanced amplification.

The paper is organized as follows. Section~\ref{Sec:MM} introduces the mathematical model of the fractionally damped unidirectionally coupled Duffing system. Section~\ref{Sec:DF} presents the diagnostic quantities and parameter-setting strategy used throughout the work. Section~\ref{Sec:RD} contains the numerical results and discussion, including the reference response of the isolated driver, the baseline and detuned coupled dynamics, and the parametric analyses in the $(F,c)$ and $(q_2,\omega_2)$ planes. Finally, Section~\ref{Sec:Conc} summarizes the main conclusions and outlines possible extensions of the present study.

\section{Mathematical Model: Fractional-Order Coupled Duffing System}\label{Sec:MM}

We consider two Duffing oscillators arranged in a unidirectionally coupled driver--receiver configuration with fractional damping. The driver is directly forced, whereas the receiver is connected to the driver through a linear coupling spring. The governing equations are
\begin{align}
&\ddot{x}(t) + \mu_1 D_t^{q_1} x(t) + \omega_1^2 x(t) + \beta x^3(t) = F \cos(\Omega t),
&\qquad 0<q_1<1, \\
&\ddot{y}(t) + \mu_2 D_t^{q_2} y(t) + \omega_2^2 y(t) + \beta y^3(t) + c (y-x) = 0,
&\qquad 0<q_2<1.
\end{align}
Here, $x(t)$ and $y(t)$ denote the displacements of the driver and receiver, respectively; $\omega_1$ and $\omega_2$ are their natural frequencies; $\beta$ is the cubic stiffness coefficient; $\mu_1$ and $\mu_2$ determine the strength of fractional damping; $F$ and $\Omega$ are the amplitude and frequency of the external forcing; and $c$ is the coupling strength.

The coupling is unidirectional: the driver dynamics do not depend on the receiver, whereas the receiver responds both to its own intrinsic dynamics and to the motion transmitted through the coupling term $c(y-x)$. This configuration is the same as that adopted in \cite{CoccoloSanjuanTR} and provides a natural framework for assessing whether fractional damping can induce an effect analogous to that previously associated with delay. At the same time, the linear spring form of the coupling keeps the interaction mechanism simple and analytically transparent.

The fractional derivative is taken in the Caputo sense,
\begin{equation}
D_t^{q} x(t) = \frac{1}{\Gamma(1-q)} \int_0^t \frac{\dot{x}(\tau)}{(t-\tau)^q} \, d\tau,
\qquad 0<q<1.
\end{equation}
Because of its nonlocal character, the Caputo derivative introduces memory into the damping term through a power-law weighting of past velocities. Fractional damping therefore provides a natural framework for examining whether continuous memory alone can reshape the resonance structure in a way analogous, at least phenomenologically, to delay-induced mechanisms.

\section{Diagnostic framework and parameter setting}\label{Sec:DF}

To quantify the strength of the fundamental resonant response, we also consider the quality factors of the driver and the receiver, defined here in the same operational sense as in Ref.~\cite{Landa}. For a generic steady-state signal $u(t)$, evaluated over the asymptotic window $[t_{\mathrm{cut}},\, t_{\mathrm{cut}}+n_T T]$ with $T=2\pi/\Omega$, we compute the sine and cosine components
\begin{align}
B_s &= \frac{2}{n_T T}\int_{t_{\mathrm{cut}}}^{t_{\mathrm{cut}}+n_T T} u(t)\sin(\Omega t)\,dt,\\
B_c &= \frac{2}{n_T T}\int_{t_{\mathrm{cut}}}^{t_{\mathrm{cut}}+n_T T} u(t)\cos(\Omega t)\,dt,
\end{align}
where $n_T$ is the number of complete forcing periods contained in the steady-state window. The corresponding quality factor is then defined as
\begin{equation}
Q[u](\Omega)=\frac{1}{\Omega}\sqrt{B_s^2+B_c^2}.
\end{equation}
Accordingly, for the driver and the receiver we define
\begin{equation}
Q_x(\Omega)=Q[x](\Omega),\qquad Q_y(\Omega)=Q[y](\Omega).
\end{equation}
In numerical practice, the integrals are evaluated over the steady-state window by means of the trapezoidal rule. In the present work, this quantity is used as a frequency-dependent indicator of the strength of the fundamental harmonic response at the forcing frequency.

We then monitor the instantaneous mechanical energies of the driver, the receiver, and the coupling spring, defined as
\begin{align}
E_1(t) &= \frac{1}{2}\dot{x}^2 + \frac{1}{2}\omega_1^2 x^2 + \frac{1}{4}\beta x^4, \\
E_2(t) &= \frac{1}{2}\dot{y}^2 + \frac{1}{2}\omega_2^2 y^2 + \frac{1}{4}\beta y^4, \\
E_c(t) &= \frac{1}{2}c (y-x)^2 .
\end{align}
Here, $E_1(t)$ and $E_2(t)$ denote the instantaneous energies of the driver and receiver oscillators, respectively, while $E_c(t)$ measures the elastic energy stored in the coupling spring. Although the system is nonconservative because of fractional dissipation, these quantities remain useful diagnostics for tracking the redistribution of kinetic, potential, and coupling-stored energy during the steady response.

To assess the direction of energy exchange through the coupling channel, we adopt the power-flow convention
\begin{equation}
P_c(t) = c (y-x)\dot{y}.
\end{equation}

Under this convention,$P_c(t)$ represents the instantaneous rate at which the coupling spring exchanges work with the receiver coordinate. Accordingly, $P_c(t)>0$ is consistent with an instantaneous transfer of energy toward the receiver, whereas $P_c(t)<0$ is consistent with energy being returned to the coupling spring. To characterize the net balance over a steady observation window, we consider the time-averaged coupling power
\begin{equation}
\langle P_c \rangle = \frac{1}{T}\int_{t_0}^{t_0+T} P_c(t)\,dt,
\end{equation}
where  $[t_0,t_0+T]$ is chosen inside the asymptotic regime after transients have decayed. This quantity provides a compact indicator of the average energetic role of the coupling channel during the periodic or nearly periodic steady response. Within the present unidirectional model, negative values of $\langle P_c \rangle$ should not be interpreted as a dynamical feedback of the receiver on the driver equation. Rather, they are consistent with the receiver--coupling subsystem releasing, on average, part of the energy previously accumulated within that part of the system through the coupling channel, according to the adopted convention.

For each forcing frequency $\Omega$, we compute the steady-state response amplitudes of the driver and receiver, denoted by $|X|$ and $|Y|$, respectively. These quantities are extracted from the asymptotic oscillatory regime and are used to construct the frequency-response curves. We also consider the phase difference between the receiver and driver responses,
\begin{equation}
\phi_{yx} = \phi_y - \phi_x,
\end{equation}
which helps distinguish frequency regions in which the two oscillators respond approximately in phase from those in which they approach an out-of-phase configuration. In combination with the amplitude and energy-flow diagnostics, this phase indicator helps identify qualitative changes in the coupled response. In addition, we define the transmissibility as
\begin{equation}
T = \frac{|Y|}{|X|},
\end{equation}
which provides a dimensionless measure of the relative amplification of the receiver with respect to the driver.

The analysis in this work is based on the combined use of these amplitude, phase, and average energy-flow indicators. Within this framework, we refer to \emph{transmitted resonance} as the conventional receiver-amplification scenario in which a local maximum of $|Y|(\Omega)$ is associated with direct excitation through the coupling spring and with a coupling-power balance that does not show a clearly storage-dominated character. In the present numerical results, this regime is typically characterized by values of $\langle P_c\rangle$ that remain close to zero, or only weakly negative, in contrast with the clearly negative values observed in the storage-dominated resonance regime. We also identify a second type of response, characterized by a local maximum of $|Y|(\Omega)$, a coherent phase relation between the oscillators, typically with $\phi_{yx}$ close to $0$ or $-\pi$, and clearly negative time-averaged coupling power, $\langle P_c \rangle < 0$. In the present paper, this regime is interpreted as a \emph{storage-dominated resonance}, in the sense that the receiver exhibits a pronounced oscillation while the average energetic balance at the coupling channel is no longer that of direct net injection from the driver. Instead, the results suggest that temporary energy accumulation and delayed release within the receiver--coupling subsystem play an important role in sustaining the observed response. This classification is intended as an operational framework for interpreting the numerical results, and should therefore be understood in conjunction with the full set of response curves, phase behavior, and energy measures presented in the next section.

To isolate the effect of fractional memory on resonance transmission, the forcing amplitude $F$, nonlinear stiffness $\beta$, driver fractional order $q_1$, driver natural frequency $\omega_1$, and driver damping coefficient $\mu_1$ are kept fixed throughout the study. In this way, the observed changes in the receiver response can be attributed to variations in the receiver memory properties and coupling-related parameters. By contrast, the receiver fractional order $q_2$, coupling strength $c$, and receiver natural frequency $\omega_2$ are systematically varied, since these quantities directly influence the balance between dissipation, coupling-mediated energy exchange, and spectral interaction between the two oscillators.

A further issue is that the effective contribution of the fractional damping term depends on frequency. If $\mu_2$ were kept fixed while varying $q_2$, the overall dissipation level would change in a trivial way, making direct comparisons across fractional orders less meaningful. To reduce this bias, we normalize the receiver damping coefficient with respect to a reference excitation frequency $\Omega^\ast$. Let $\mu_2$ denote the coefficient multiplying the fractional damping term $D_t^{q_2}y(t)$. Since its effective contribution scales as $\Omega^{q_2}$, we impose that the effective damping at the prescribed reference frequency $\Omega^\ast$ remain constant for all $q_2$, namely
\begin{equation}
\mu_2(q_2)(\Omega^\ast)^{q_2} = K,
\end{equation}
where $K$ is a constant determined from a reference choice $(q_2^{\mathrm{ref}},\mu_2^{\mathrm{ref}})$. Solving for $\mu_2(q_2)$ gives
\begin{equation}
\mu_2(q_2) = \mu_2^{\mathrm{ref}}(\Omega^\ast)^{q_2^{\mathrm{ref}}-q_2}.
\end{equation}
In the simulations reported below, we choose $\Omega^\ast = 2$, which is close to the main resonance frequency of the driver, together with $q_2^{\mathrm{ref}}=0.1$ and $\mu_2^{\mathrm{ref}}=0.5$. This yields
\begin{equation}
\mu_2(q_2)=0.5\times 2^{0.1-q_2}.
\end{equation}
With this normalization, comparisons across different values of $q_2$ are performed under approximately equivalent effective damping conditions at the reference frequency, so that the resulting differences can be more reasonably attributed to memory effects rather than to a trivial rescaling of dissipation strength.

\section{Results and discussion}\label{Sec:RD}

In this section, we examine how fractional memory and coupling shape the resonant response of the system. We begin with the isolated driver in order to define a reference state, then turn to the baseline and detuned coupled dynamics, and finally analyze the global organization of the response through parametric maps in the $(F,c)$ and $(q_2,\omega_2)$ planes.

\subsection{Reference response of the isolated driver}

We begin by examining the driver oscillator in isolation in order to select a representative fractional order for the forcing subsystem. For the parameter set considered here, decreasing the driver fractional order enhances the resonant response of the isolated fractional Duffing oscillator, leading to larger amplitude, larger average energy, and a sharper quality-factor peak. On this basis, we adopt $q_1=0.1$ in the coupled-system analysis, since this value provides the strongest driver response among the cases considered and therefore offers a suitable reference for studying resonance transmission to the receiver.

\begin{figure}[!t]
\centering
\begin{subfigure}{0.32\textwidth}
    \centering
    \begin{overpic}[width=\linewidth]{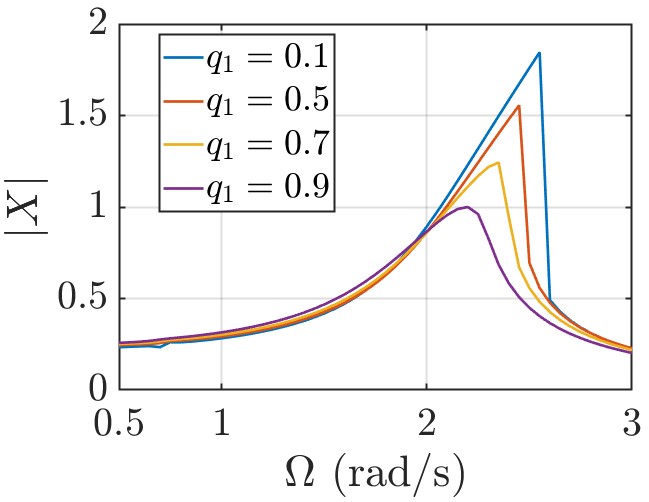}
        \put(-2,4){\Large\bfseries (a)}
    \end{overpic}
    \phantomcaption\label{fig:Ax}
\end{subfigure}
\hfill
\begin{subfigure}{0.32\textwidth}
    \centering
    \begin{overpic}[width=\linewidth]{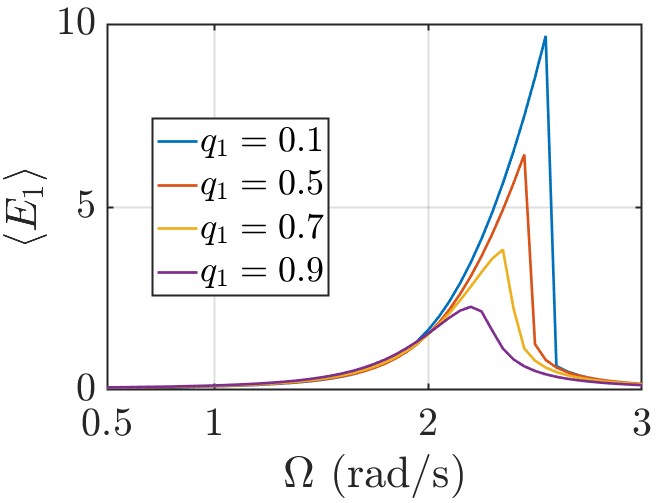}
        \put(-4,4){\Large\bfseries (b)}
    \end{overpic}
    \phantomcaption\label{fig:E1}
\end{subfigure}
\hfill
\begin{subfigure}{0.32\textwidth}
    \centering
    \begin{overpic}[width=\linewidth]{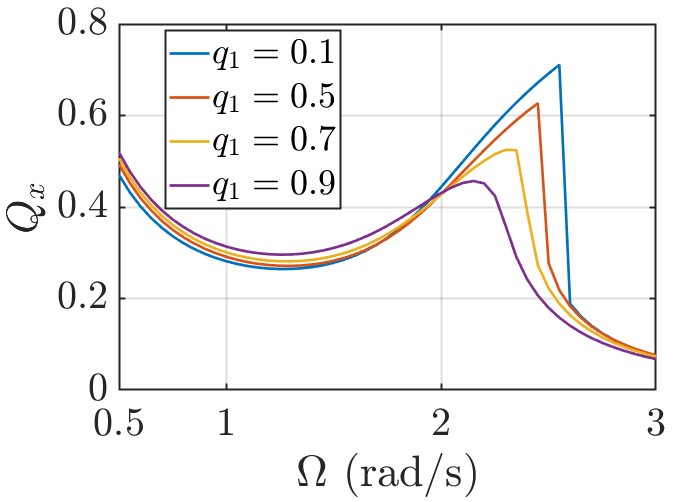}
        \put(-2,4){\Large\bfseries (c)}
    \end{overpic}
    \phantomcaption\label{fig:Qx}
\end{subfigure}
\caption{\justifying Frequency-response characteristics of the isolated fractional Duffing driver for several values of the fractional order $q_1$: (a) steady-state amplitude $|X|$, (b) average driver energy $E_1$, and (c) quality factor $Q_x$, all plotted as functions of the forcing frequency $\Omega$. Decreasing $q_1$ enhances the resonance peak in all three diagnostics, leading to stronger amplitude amplification, larger stored energy, and a sharper resonant response. This behavior motivates the choice $q_1=0.1$ for the coupled-system study. Parameters: $\omega_1=2$, $\beta=1$, $F=1$, and $\mu_1=0.5$.}
\label{fig:main_results}
\end{figure}

Figure~\ref{fig:main_results} shows the frequency-response characteristics of the isolated driver for several values of $q_1$, with $\omega_1=2$, $\beta=1$, $F=1$, and $\mu_1=0.5$. As $q_1$ decreases, the resonance peak becomes more pronounced in all three diagnostics. The amplitude $|X|$ increases, the average energy $E_1$ reaches larger values, and the quality factor $Q_x$ develops a sharper maximum around the main resonance region. This behavior indicates that reducing the fractional order weakens the effective dissipative action near resonance and allows a more coherent oscillatory response to develop. In particular, the case $q_1=0.1$ yields the largest amplitude and the sharpest resonance peak, and is therefore used throughout the coupled-system analysis below.

\subsection{Baseline coupled response: transmitted and storage-dominated resonance}

We now turn to the unidirectionally coupled system governed by Eqs.~(1) and (2), and analyze how the receiver response is shaped by the coupling spring when continuous memory is present in both oscillators. For the reference parameter set, the frequency-response curves reveal two distinct maxima in the receiver amplitude, indicating that the coupled dynamics cannot be reduced to a single transmitted resonance peak. This baseline parameter set was selected as a representative operating regime in which the coupled system clearly exhibits both the conventional transmitted-resonance peak and the higher-frequency storage-supported response, thereby providing a convenient reference configuration for the subsequent detuning and parametric analyses.

\begin{figure}[!t]
    \centering
    \includegraphics[width=1.0\linewidth]{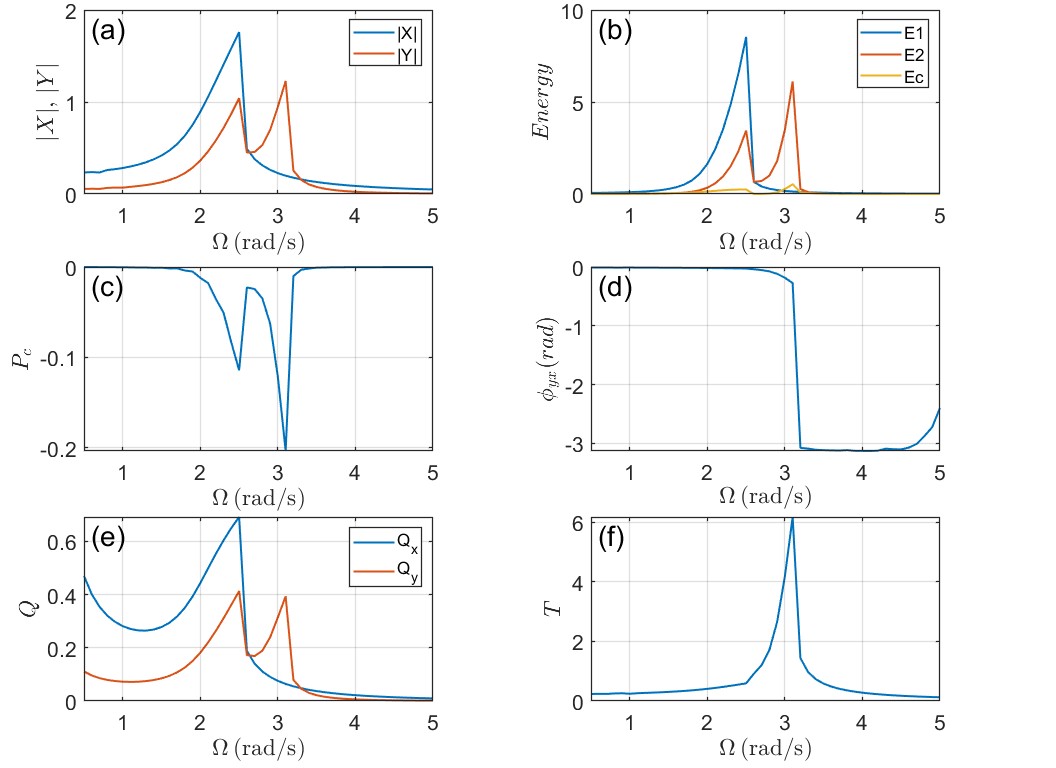}
\caption{\justifying Baseline frequency-response analysis of the coupled system for $\omega_1=2$, $\omega_2=2.5$, $\beta=1$, $F=1$, $c=2$, $\mu_1=0.5$, $\mu_2=0.5$, $q_1=0.1$, and $q_2=0.1$: (a) steady-state amplitudes $|X|$ and $|Y|$, (b) average energies $E_1$, $E_2$, and $E_c$, (c) time-averaged coupling power $\langle P_c\rangle$, (d) phase difference $\phi_{yx}$, (e) quality factors $Q_x$ and $Q_y$, and (f) transmissibility $T=|Y|/|X|$, all as functions of the forcing frequency $\Omega$. The figure highlights the two distinct receiver response regions and their associated changes in energy transfer, phase relation, and relative amplification.}
\label{fig:baseline_response}
\end{figure}

Figure~\ref{fig:baseline_response}(a) compares the steady-state amplitudes of the driver and receiver as functions of the forcing frequency. The first receiver peak appears close to the main resonance region of the driver and corresponds to the standard situation in which the oscillatory response is transmitted through the coupling spring. A second peak emerges at higher forcing frequency, where the receiver again develops a pronounced oscillation despite the driver being already beyond its main resonance region. This second response suggests the activation of an additional coupled dynamical mechanism. In the numerical results, it becomes more visible when the receiver fractional order is small, which indicates that fractional memory favors the persistence of this higher-frequency response.

The energetic origin of these two response regions is clarified in Fig.~\ref{fig:baseline_response}(b), where the average energies of the driver, the receiver, and the coupling spring are shown. The driver energy $E_1$ reaches its dominant maximum near the main forced-resonance region, while the receiver energy $E_2$ displays two local maxima aligned with the two peaks observed in the amplitude response. At the same time, the energy stored in the coupling spring, $E_c$, increases significantly in both regions, showing that the elastic interaction between the oscillators becomes particularly active there. These results indicate that the second receiver peak is not a minor residual feature, but is accompanied by a clear reorganization of the energy distribution within the coupled system.

A key feature of the second response region appears in Fig.~2(c), which shows the time-averaged coupling power $\langle P_c\rangle$. In the lower-frequency response region, $\langle P_c\rangle$ remains close to zero, indicating that the receiver response is not associated with a strong net release of energy through the coupling spring. By contrast, near the higher-frequency peak, $\langle P_c\rangle$ becomes clearly negative. According to the convention adopted in Sec.~III, this indicates that the receiver--coupling subsystem releases, on average, part of the energy previously accumulated within that part of the system through the coupling spring.

This interpretation is reinforced by the phase behavior displayed in Fig.~\ref{fig:baseline_response}(d). As the forcing frequency crosses the second response region, the phase difference $\phi_{yx}$ undergoes a sharp variation, evolving from a nearly in-phase configuration toward a state approaching anti-phase motion. This transition indicates that the second receiver peak corresponds to a qualitatively different oscillatory regime, rather than to a simple high-frequency tail of the main resonance.

Further information is provided by the quality factors shown in Fig.~\ref{fig:baseline_response}(e). As expected, $Q_x$ exhibits its dominant peak near the main resonance of the driven oscillator. The receiver quality factor $Q_y$, however, displays two distinct maxima, reflecting the existence of two frequency intervals in which the receiver develops a relatively coherent amplified response. In particular, the second peak in $Q_y$ shows that the higher-frequency receiver response remains structured and localized, even though the average energy-transfer balance differs from that of the lower-frequency peak.

The transmissibility, $T=|Y|/|X|$, is shown in Fig.~\ref{fig:baseline_response}(f). It exhibits a pronounced maximum near the second response region, indicating that the receiver amplitude becomes large relative to that of the driver. This does not imply that the receiver always stores more total energy than the driver, but rather that the relative oscillatory amplification is particularly strong there. Together with the phase variation and the sign change of $\langle P_c\rangle$, the transmissibility peak supports the conclusion that the second response region is sustained by a mechanism more complex than direct resonance transmission alone.

Taken together, the baseline coupled response shows that fractional damping modifies resonance transmission in a nontrivial way. Near the primary response region, the receiver behavior remains closer to the conventional transmitted-resonance picture. At higher forcing frequency, however, the receiver develops a second pronounced response characterized by large relative amplitude, a marked phase transition, and clearly negative time-averaged coupling power. This combination of features indicates that the higher-frequency response is influenced not only by direct transmission from the driver, but also by temporary energy accumulation and delayed release within the receiver--coupling subsystem. In this sense, fractional memory reshapes the energetic pathways through which the receiver response is amplified.

\subsection{Detuning-induced amplification and superposed resonance}

We next examine how this structure changes when the receiver natural frequency is detuned with respect to that of the driver. This detuning modifies the spectral interaction between the lower-frequency transmitted response and the higher-frequency coupled response, and can therefore lead to a substantial enhancement of the receiver oscillation. Although this effect is conceptually related to the resonance-transmission and resonance-superposition phenomena previously reported in delayed Duffing-type settings \cite{CoccoloSanjuanDelay,CoccoloSanjuanTR}, the underlying memory mechanism is different here, since it originates from fractional damping rather than from discrete delay.

\begin{figure}[!t]
    \centering
    \includegraphics[width=0.9\linewidth]{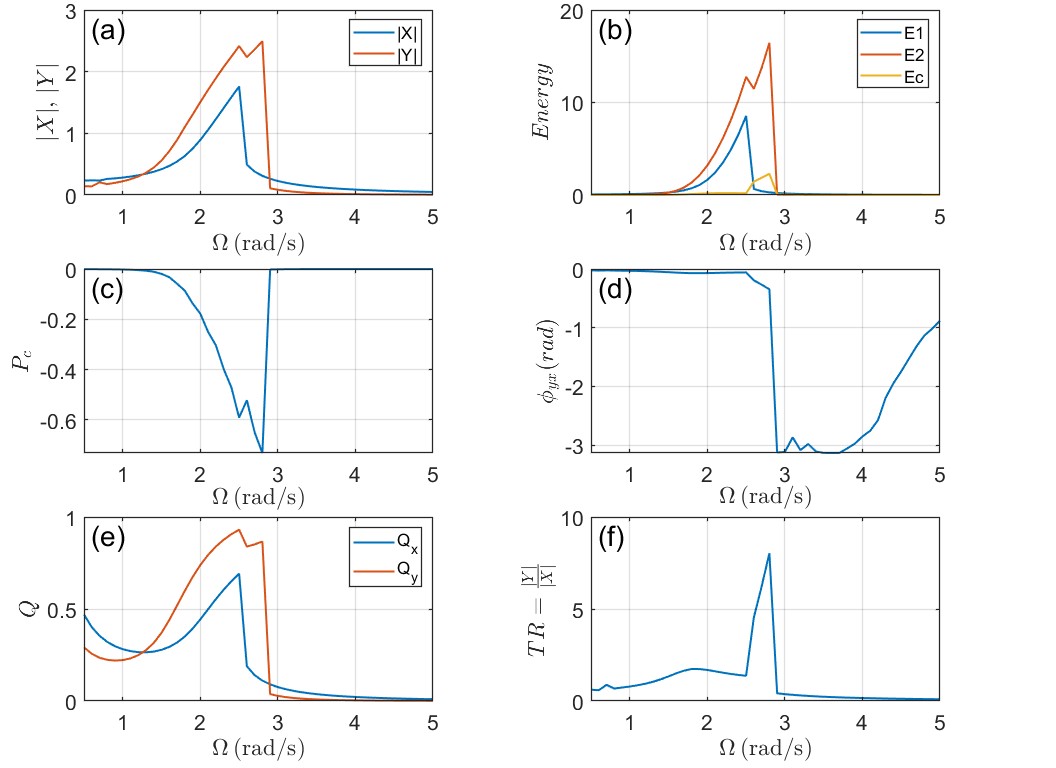}
\caption{\justifying Detuned coupled response for $\omega_2<\omega_1$: (a) amplitudes $|X|$ and $|Y|$, (b) average energies $E_1$, $E_2$, and $E_c$, (c) time-averaged coupling power $\langle P_c\rangle$, (d) phase difference $\phi_{yx}$, (e) quality factors $Q_x$ and $Q_y$, and (f) transmissibility $T$, all plotted versus the forcing frequency $\Omega$. Detuning the receiver natural frequency enhances the receiver response and promotes a stronger interaction between direct transmission and storage-mediated amplification. Parameters: $\omega_1=2$, $\omega_2=1$, $\beta=1$, $F=1$, $c=2$, $\mu_1=0.5$, $\mu_2=0.5$, $q_1=0.1$, and $q_2=0.1$.}
    \label{fig:combined}
\end{figure}

Figure~\ref{fig:combined} presents a global view of the detuned case. Decreasing $\omega_2$ below $\omega_1$ significantly strengthens the receiver response. In particular, the receiver amplitude, receiver energy, coupling-spring energy, and transmissibility all increase in the detuned regime, and for suitable forcing frequencies the receiver response becomes comparable to, or even larger than, that of the driver. At the same time, the phase behavior and the time-averaged coupling power show that this enhancement is not explained by direct transmission alone. Instead, the results support the interpretation that detuning promotes a constructive interaction between the lower-frequency transmitted response and the higher-frequency coupled response. Combined with the storage effects associated with fractional damping, this interaction gives rise to a markedly amplified receiver oscillation. In this sense, detuning acts as an additional control mechanism for resonance localization and amplification in the fractionally damped coupled system.

\subsection{Parametric maps in the $(F,c)$ plane}

To complement the frequency-response analysis, we now examine how the different response regimes are organized across parameter space. While the one-parameter curves reveal the local mechanisms underlying the baseline and detuned responses, they do not fully show how forcing intensity, coupling strength, receiver fractional order, and receiver natural frequency interact to shape the global dynamics. We therefore perform a systematic parametric study using two sets of maps: first in the $(F,c)$ plane for several representative values of $q_2$, and then in the $(q_2,\omega_2)$ plane in order to analyze how fractional memory and detuning jointly tune the receiver response.

\begin{figure}[!t]
    \centering
    \includegraphics[width=17.0cm,clip=true]{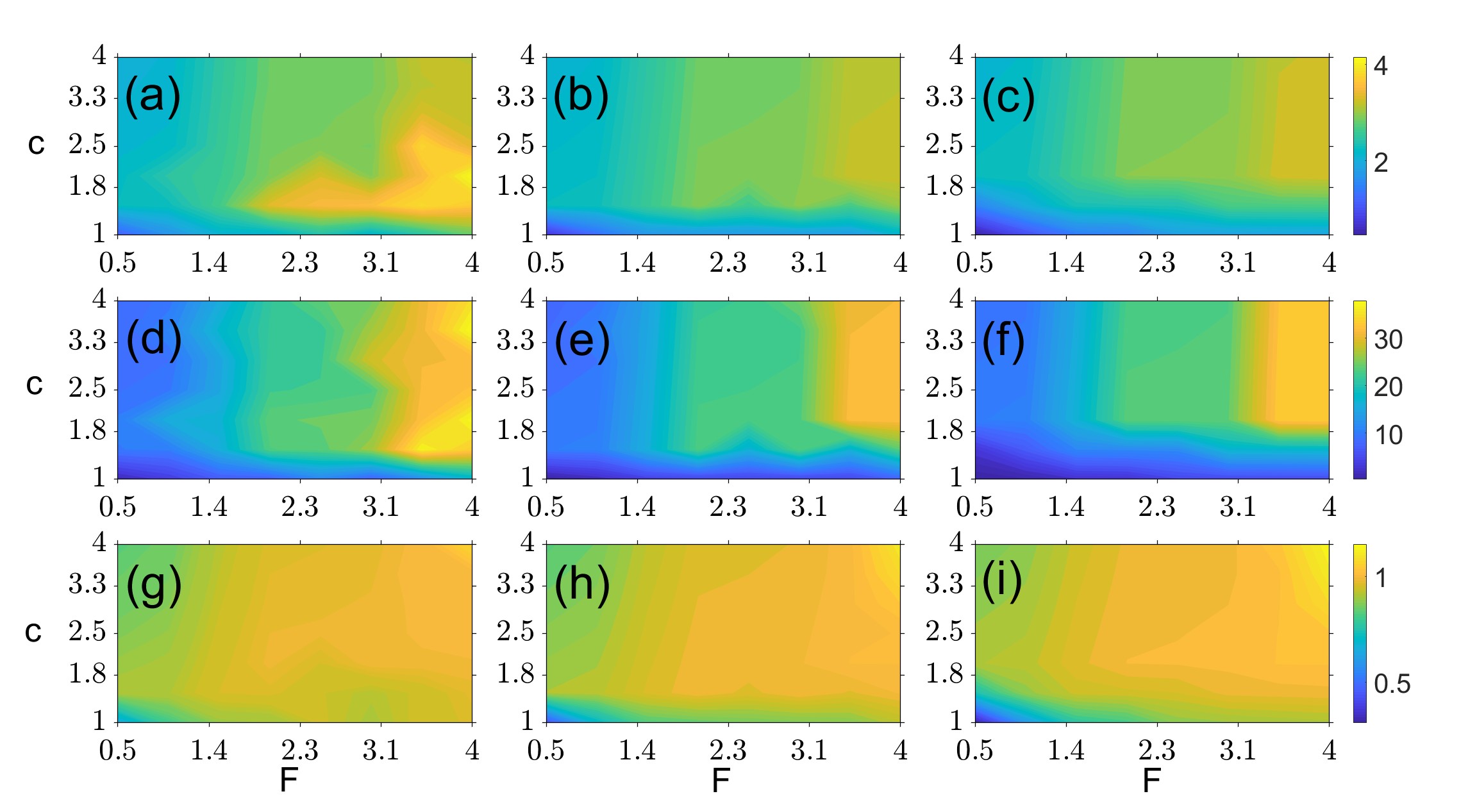}
\caption{\justifying Maps in the $(F,c)$ plane for representative values of the fractional order $q_2$. Panels (a)--(c) show the maximum steady-state receiver amplitude $|Y|$ for $q_2=0.1$, $0.5$, and $0.9$, respectively; panels (d)--(f) show the corresponding time-averaged receiver energy; and panels (g)--(i) show the receiver quality factor $Q_y$. Taken together, the nine panels show that strong receiver amplification emerges above a forcing threshold and is most sharply localized for small fractional order. The coincidence between the amplitude and energy corridors, together with the narrow high-$Q_y$ regions, indicates that fractional memory favors selective and coherent resonance localization.}
\label{fig:Fc_response_maps}
\end{figure}

The maps in Fig.~\ref{fig:Fc_response_maps}(a)--(f) display the receiver response in the $(F,c)$ plane for $q_2=\{0.1,0.5,0.9\}$. More specifically, Fig.~\ref{fig:Fc_response_maps}(a)--(c) shows the maximum receiver amplitude $|Y|$, whereas Fig.~\ref{fig:Fc_response_maps}(d)--(f) shows the corresponding time-averaged receiver energy. Both diagnostics reveal a clear activation threshold with respect to the forcing amplitude $F$. For weak forcing, the receiver remains in a low-response state over the full coupling range, indicating that the oscillatory motion transmitted through the coupling spring is insufficient to generate significant amplification. Once the forcing exceeds a certain level, however, localized high-response corridors emerge in parameter space.

These corridors become particularly sharp for small fractional order. For $q_2=0.1$, shown in Fig.~\ref{fig:Fc_response_maps}(a) and Fig.~\ref{fig:Fc_response_maps}(d), the receiver develops narrow regions of strong amplification, showing that low fractional order favors highly selective resonance activation. As $q_2$ increases to $0.5$ and $0.9$, shown in Fig.~\ref{fig:Fc_response_maps}(b), (c), (e), and (f), the response becomes less localized: the high-response regions broaden, but their peak intensity decreases. This indicates that increasing the fractional order redistributes the response over a wider portion of parameter space while reducing the maximum amplification sustained by the receiver.

The quality-factor maps shown in Fig.~\ref{fig:Fc_response_maps}(g)--(i) provide further evidence of this trend. For small fractional order, the receiver exhibits large values of $Q_y$ concentrated along the high-amplitude corridors, indicating a narrow and coherent response. For larger $q_2$, the quality factor decreases and the high-$Q_y$ regions become more diffuse, showing that the resonant response becomes less sharply localized as the memory effect becomes less favorable to selective amplification.

\begin{figure}[!t]
    \centering
    \includegraphics[width=17.0cm,clip=true]{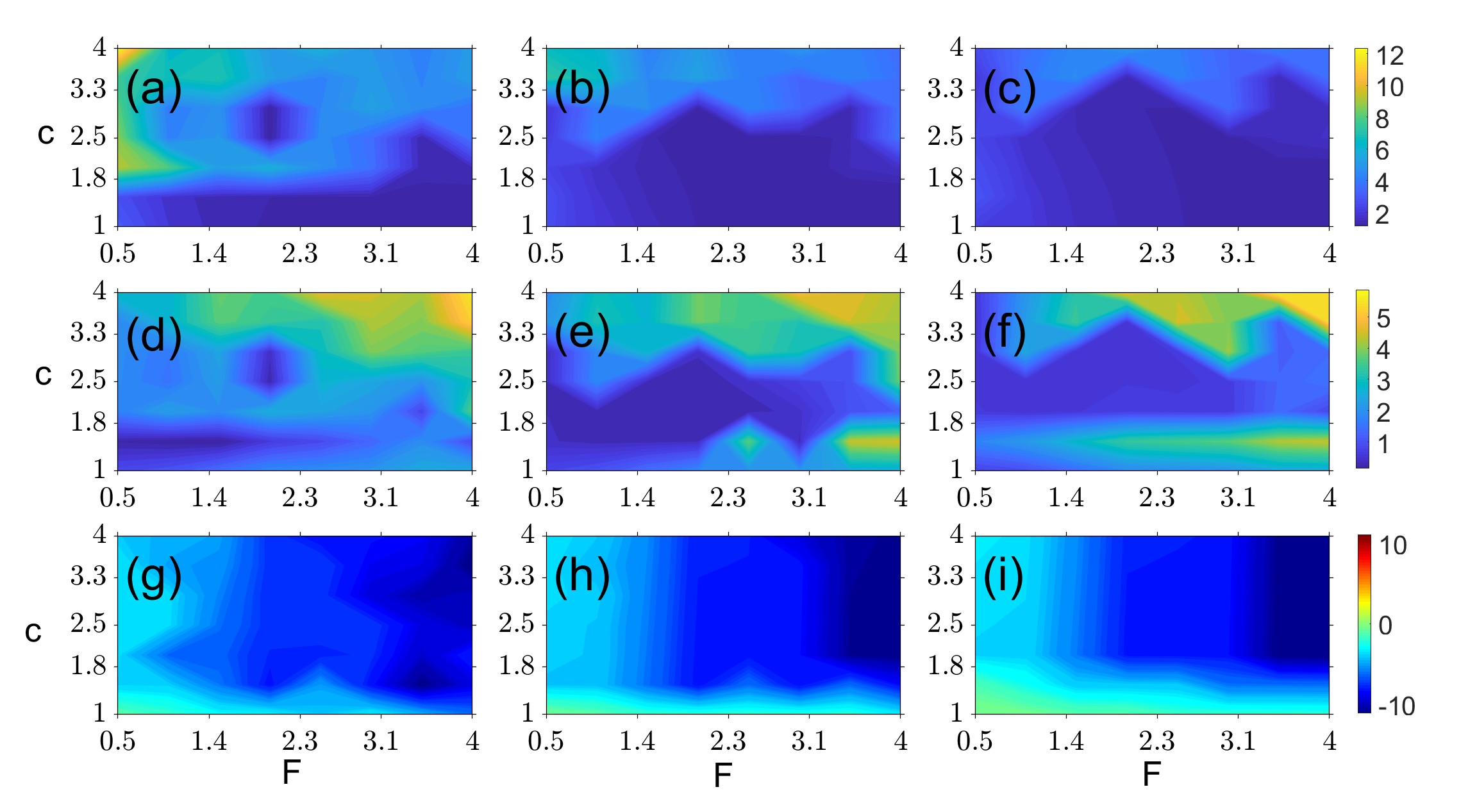}
\caption{\justifying Maps in the $(F,c)$ plane for representative values of the fractional order $q_2$. Panels (a)--(c) show the transmissibility $T=|Y|/|X|$ for $q_2=0.1$, $0.5$, and $0.9$, respectively; panels (d)--(f) show the energy stored in the coupling spring $E_c$; and panels (g)--(i) show the time-averaged coupling power $\langle P_c\rangle$. Together, these panels show that the strongest receiver response cannot be interpreted in terms of direct transmission alone. Regions of strong amplification are accompanied by enhanced energy storage in the coupling spring, while extended domains of negative $\langle P_c\rangle$ indicate release of previously stored energy from the receiver--coupling subsystem, which is characteristic of storage-dominated resonance.}
\label{fig:Fc_energy_maps}
\end{figure}

A different perspective is provided by the transmissibility and coupling-spring diagnostics shown in Fig.~\ref{fig:Fc_energy_maps}(a)--(i). The transmissibility maps in Fig.~\ref{fig:Fc_energy_maps}(a)--(c) show where the receiver becomes strongly amplified relative to the driver. For $q_2=0.1$, shown in Fig.~\ref{fig:Fc_energy_maps}(a), large values of $T$ occupy well-defined corridors, indicating efficient transmission of oscillatory motion from the driver to the receiver. As the fractional order increases, the corresponding regions shown in Fig.~\ref{fig:Fc_energy_maps}(b) and (c) become less localized and the transmissibility weakens over most of the parameter plane.

This reduction in transmissibility does not, however, eliminate the high receiver amplitudes found in Fig.~\ref{fig:Fc_response_maps}(a)--(c). Strong receiver oscillations may persist even where $T$ is only moderate, which shows that direct transmission alone is not sufficient to explain the response. The coupling-spring energy shown in Fig.~\ref{fig:Fc_energy_maps}(d)--(f) clarifies this point: in the same parameter regions where the receiver becomes strongly excited, the energy stored in the coupling spring also increases significantly. This indicates that the coupling spring acts as a temporary energy reservoir rather than as a purely passive transmission element.

The mean coupling power shown in Fig.~\ref{fig:Fc_energy_maps}(g)--(i) adds the final piece of the picture. Large regions of negative $\langle P_c\rangle$ appear in the same broad response domains where the receiver remains strongly excited. According to the convention adopted in this work, this means that part of the energy previously accumulated in the receiver--coupling subsystem is released through the coupling spring, rather than the response being sustained exclusively by direct net injection from the driver. The combined interpretation of $T$, $E_c$, and $\langle P_c\rangle$ therefore shows that the global response is governed by two intertwined mechanisms: direct transmitted resonance and storage-dominated resonance. In this sense, the $(F,c)$ maps confirm from a broader parameter perspective the energetic interpretation already identified in the frequency-response curves.

\begin{figure}[!t]
    \centering
    \includegraphics[width=1\linewidth]{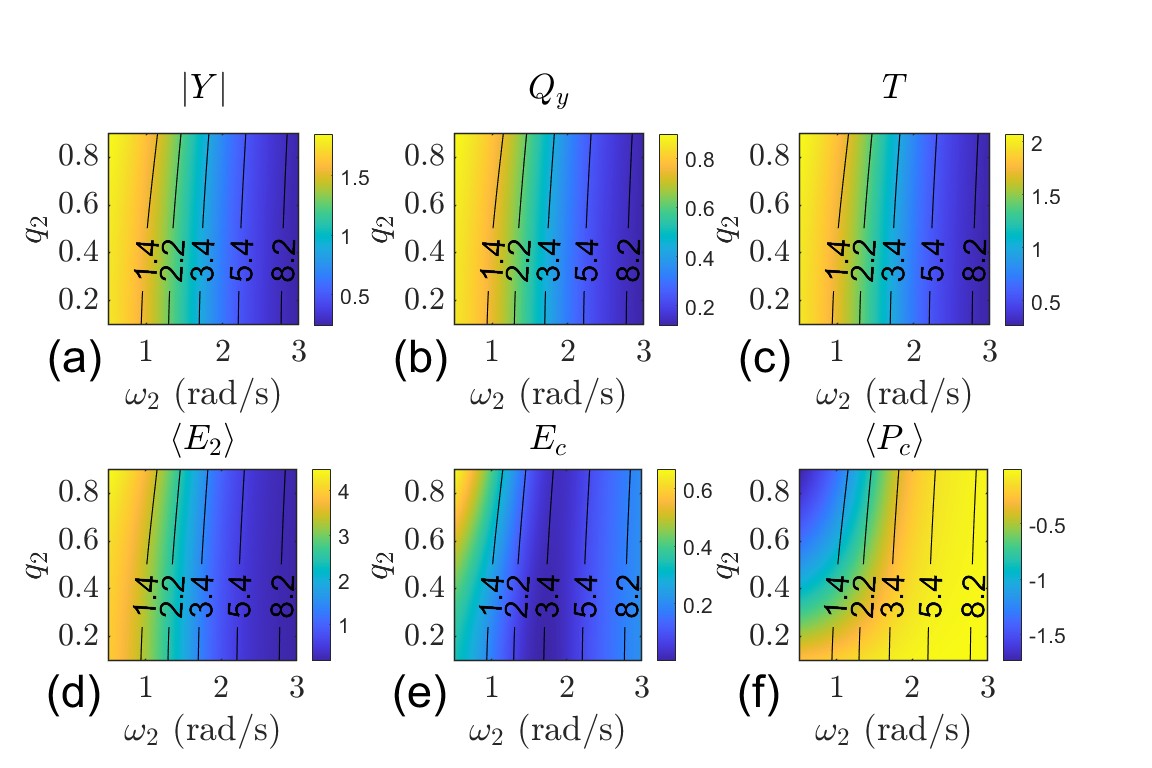}
\caption{\justifying Maps in the $(q_2,\omega_2)$ plane showing, from left to right and top to bottom, the receiver amplitude $|Y|$, receiver quality factor $Q_y$, transmissibility $T$, time-averaged receiver energy, energy stored in the coupling spring, and time-averaged coupling power $\langle P_c\rangle$. Superimposed curves denote selected contour lines of the heuristic effective coefficient $k_{\mathrm{eff}}$, defined by $k_{\mathrm{eff}}(q_2,\omega_2)=K$, where the numerical labels indicate the corresponding constant value $K$. These contours are included only as a qualitative guide for comparing the smooth organization of the maps with the variation of the effective restoring term across the $(q_2,\omega_2)$ plane.}
    \label{fig:q2w2_maps}
\end{figure}

\subsection{Joint influence of fractional order and detuning in the $(q_2,\omega_2)$ plane}

After establishing the main response organization in the $(F,c)$ plane, we turn to the $(q_2,\omega_2)$ plane in order to isolate the joint role of fractional memory and receiver detuning under otherwise fixed coupling conditions. Figure~\ref{fig:q2w2_maps} shows a set of maps in the $(q_2,\omega_2)$ plane displaying the receiver amplitude, receiver quality factor, transmissibility, time-averaged receiver energy, energy stored in the coupling spring, and time-averaged coupling power. Taken together, these maps provide a synthetic view of how fractional memory and detuning combine to reshape the receiver response.

In contrast to the sharper corridors observed in the $(F,c)$ plane, the dependence on $(q_2,\omega_2)$ is smoother and more gradual. The amplitude, quality-factor, transmissibility, and receiver-energy maps show a systematic variation across the plane, indicating that the response is continuously tuned by the combined action of memory and detuning rather than organized into strongly localized narrow ridges. In particular, lower values of $\omega_2$ and, to a lesser extent, lower fractional orders tend to favor stronger receiver response, larger quality factor, and larger transmissibility.

At a heuristic level, part of this smooth organization can be related to the frequency-dependent contribution of the fractional damping term through an effective linear coefficient of the form
\begin{equation}
k_{\mathrm{eff}} \approx \omega_2^2 + \mu_2 \Omega^{q_2}\cos\left(\frac{q_2\pi}{2}\right).
\end{equation}
This expression is not introduced as a strict reduced model, but only as a qualitative guide. To facilitate the interpretation of the maps in the $(q_2,\omega_2)$ plane, we superimpose selected contour lines of $k_{\mathrm{eff}}$, defined by the level-set condition
\[
k_{\mathrm{eff}}(q_2,\omega_2)=K,
\]
for representative constant values $K$. Using the damping normalization adopted in this work,
\[
\mu_2(q_2)=0.5\,2^{0.1-q_2},
\]
together with the reference frequency $\Omega_\ast=2$, these contour lines are computed directly from the implicit relation
\[
\omega_2^2 + \mu_2(q_2)\,\Omega_\ast^{q_2}\cos\!\left(\frac{\pi q_2}{2}\right)=K.
\]
The numerical labels attached to the curves indicate the corresponding constant value of $k_{\mathrm{eff}}$.

The maps of $|Y|$, $Q_y$, $T$, and the time-averaged receiver energy are broadly consistent with an anticorrelated trend with $k_{\mathrm{eff}}$, in the sense that stronger response is generally found in regions where this effective coefficient is smaller. This is physically reasonable, since a smaller effective restoring term makes the receiver more easily excitable by the transmitted forcing. At the same time, not all observables follow the same trend. The energy stored in the coupling spring exhibits a more structured and non-monotonic dependence on $(q_2,\omega_2)$, which indicates that the storage process is governed by a more delicate balance between receiver dynamics, detuning, and coupling. Likewise, the map of $\langle P_c\rangle$ does not correlate in a simple one-to-one way with $k_{\mathrm{eff}}$, showing that the average coupling-power balance cannot be reduced to an effective-stiffness argument alone. This is consistent with the fact that $\langle P_c\rangle$ also depends on relative phase organization and on the detailed energetic exchange between the receiver and the coupling channel.

Even so, the $\langle P_c\rangle$ map remains important because it identifies broad parameter regions in which the average coupling-power balance becomes weakly negative, supporting the interpretation that the amplified receiver response is not governed solely by direct transmission. Under suitable detuning conditions, this storage-supported response evolves into a stronger amplified regime in which the internally sustained oscillation of the receiver becomes favorably aligned with the transmitted forcing, and the receiver response may become comparable to or larger than that of the driver. This is the regime that we refer to as superposed resonance.

Overall, the parametric maps show that the receiver fractional order acts as a tuning parameter with a dual role. On the one hand, it modifies the effective dissipative action and therefore influences the overall response level and sharpness. On the other hand, together with the receiver natural frequency, it continuously shifts the balance between direct transmission, temporary energy storage, and amplified response. For applications in which response localization or enhanced oscillatory amplification is desirable, such as vibration control or energy harvesting, these results suggest that low fractional order can be especially advantageous when combined with suitable receiver detuning and coupling conditions. 

\section{Conclusion}\label{Sec:Conc}

In this work, we investigated resonance transmission in a system of unidirectionally coupled Duffing oscillators with fractional damping, focusing on how memory effects in the receiver modify amplitude amplification, energy redistribution, and the overall structure of the coupled response. The central novelty of the paper is that the receiver dynamics cannot always be interpreted in terms of direct transmitted resonance alone. In addition to the standard transmitted-resonance picture, in which the receiver response is associated with direct energy exchange through the coupling spring, the numerical results reveal a second regime in which the receiver still displays a pronounced oscillatory response while the time-averaged coupling power becomes clearly negative under the adopted convention. Within the present unidirectional model, this does not imply feedback on the driver, but instead indicates that part of the energy previously accumulated in the receiver--coupling subsystem is released through the coupling spring. This provides the basis for identifying a storage-dominated resonance mechanism in fractionally damped coupled oscillators.

To establish this picture, we combined frequency-response curves with energy, phase, transmissibility, quality-factor, and time-averaged coupling-power diagnostics. First, the isolated driver analysis showed that choosing a small driver fractional order enhances the resonant response and provides a suitable reference state for the coupled-system study. We then showed that, for the baseline coupled configuration, the receiver develops two distinct response peaks. The first remains closer to the conventional transmitted-resonance scenario, whereas the second is associated with a marked phase transition, enhanced relative receiver amplification, and clearly negative time-averaged coupling power. Taken together, these signatures show that fractional damping does not simply attenuate the motion but actively reshapes the energetic pathways through which the receiver response is sustained.

A second main result is that detuning the receiver natural frequency strongly enhances the coupled response. In the detuned regime, the interaction between the lower-frequency transmitted response and the higher-frequency coupled response becomes more effective, leading to a markedly amplified receiver oscillation. This amplified regime is interpreted here as superposed resonance. In this sense, the paper shows that the receiver response can be controlled not only through the coupling strength and forcing conditions, but also through the interplay between fractional memory and spectral detuning.

The parametric maps further demonstrate that the receiver fractional order $q_2$, the coupling strength $c$, and the receiver natural frequency $\omega_2$ act as effective control parameters for the response level, its sharpness, and the balance between transmission and storage. In particular, low fractional order promotes sharply localized high-response regions in the $(F,c)$ plane, whereas the joint variation of $q_2$ and $\omega_2$ produces a smoother but systematic reorganization of the response. Part of this behavior can be understood heuristically through an effective linear coefficient, for which a stronger response is broadly associated with a lower effective restoring term. At the same time, the coupling-spring energy and especially the time-averaged coupling power show that the energetic balance cannot be reduced to an effective-stiffness argument alone, since temporary storage and phase-dependent exchange remain essential ingredients of the amplified response.

Overall, the results show that nonlocal dissipation can play a constructive role in coupled nonlinear oscillators by modifying how energy is transmitted, stored, and released. Beyond the specific Duffing configuration considered here, this suggests that fractional damping may offer a useful framework for controlling resonance localization and amplification in more general coupled oscillatory systems. Future work may extend this approach to bidirectionally coupled or networked oscillators, examine alternative fractional constitutive laws, and explore possible experimental realizations of the mechanisms identified here.

\section{Acknowledgments}

This work was supported by the Spanish State Research Agency (AEI) and the European Regional Development Fund (ERDF,EU) under Project No. PID2023-148160NB-I00 (MCIN/AEI/10.13039/501100011033).

\end{document}